\documentstyle[aps,prl,multicol,amssymb,epsf]{revtex}

\title{ Multi-Particle Entanglement Purification Protocols} 

\author{M. Murao$^1$, M.B. Plenio$^1$, S. Popescu$^{2,3}$,
V. Vedral$^1$ and P.L. Knight$^1$} 

\address{$^1$Optics Section, Blackett Laboratory, Imperial College,
London SW7 2BZ, UK\\
$^2$Isaac Newton Institute for Mathematical
Sciences, Cambridge, CB3 0EH, UK\\%
$^3$BRIMS, Hewlett-Packard Laboratories, Stoke
Gifford, Bristol BS12 6QZ, UK}

\date{\today}
\begin{document}
\draft
\maketitle
\begin{abstract}
Purification schemes for multi-particle entangled states cannot be
treated as straightforward extensions of those for two particles
because of the lack of symmetry they possess.  We propose purification
protocols for a wide range of mixed entangled states of many
particles.  These are useful for understanding entanglement, and will
be of practical significance in multi-user cryptographic schemes or
distributed quantum computation and communication.  We show that
operating locally on multi-particle entangled states directly is more
efficient than relying on two-particle purification.
\end{abstract}


\begin{multicols}{2}

Entanglement is of central importance for quantum computation
\cite{Barenco}, quantum teleportation \cite{Bennett0}, and certain
types of quantum cryptography \cite{Ekert}.  Without entangled states,
quantum computation and communication would be no more efficient than
their classical counterparts. For two particles, the maximally
entangled states are the Bell diagonal states $\left \vert \phi^\pm
\right \rangle=\frac{1}{\sqrt{2}} \left( \left \vert 0 0 \right
\rangle \pm \left \vert 1 1 \right \rangle \right)$, $\left \vert
\psi^\pm \right \rangle=\frac{1}{\sqrt{2}} \left( \left \vert 0 1
\right \rangle \pm \left \vert 1 0 \right \rangle \right)$, and all
other locally unitarily equivalent ones, where the state for each
particle is written in the qubit ($\left \vert 0 \right \rangle$,
$\left \vert 1 \right \rangle$) basis. For many spin 1/2 particles,
the maximally entangled states are
\begin{eqnarray}
\left \vert \phi^\pm\right \rangle=\frac{1}{\sqrt{2}} 
\left( 
\left \vert 0 0 \cdot \cdot \cdot 0 \right \rangle
\pm
\left \vert 1 1 \cdot \cdot \cdot 1 \right \rangle
\right),
\end{eqnarray}
as well as those which are locally unitarily equivalent; for three
particles, these are called GHZ states \cite{Greenberger}.
Unfortunately entangled states turn into mixed states due to the
dissipative effects of the environment, and this is one of the main
obstacles for the practical realization of quantum computation and
entanglement based quantum cryptography.  The environment does not
always destroy entanglement completely.  Mixed states resulting from
interaction with the environment may still contain some residual
entanglement \cite{Bennett1}. The task is then to ``purify'' this
residual entanglement with the aim of obtaining maximally entangled
states.  These purification procedures use only local operations and
classical communication \cite{Bennett,Gisin,Deutsch} Related to this,
various quantitative measures of entanglement for mixed states have
been proposed \cite{Vedral1,Vedral2,Vedral3}.  Popescu and Rohrlich
\cite{Popescu2} have proven, using arguments based on purification
procedures \cite{Bennett,Gisin,Deutsch}, that the von Neumann entropy
is a unique measure of entanglement for pure bipartite states.  

These measures can give upper bounds on the efficiency with which one
can purify an initial ensemble of partially entangled states.
Disentangled states, for two particles they are of the form $\sum p_i
\rho^{1}_i \otimes \rho^{2}_i$ where $\rho^1$ and $\rho^2$ are the
local density matrices \cite{Horodecki}, cannot be purified.  For many
particles the generalisation is not unique.  One can define
disentangled states as these being of the form $\sum p_i \rho^{1}_i
\otimes \cdots \otimes \rho^{N}_i$ or as those states from which one
cannot purify using local operations a maximally entangled state.
This alternative definition gives the investigation of multi-particle
purification procedures a fundamental importance in the understanding
of entanglement.  The two definitions might proved to be  equivalent,
but this question remains open at present.

Several purification protocols have been proposed
\cite{Bennett,Gisin,Deutsch} for the purification of two-particle
entangled states.  For two particles, the singlet state ($\left \vert
\phi^- \right \rangle$), which is totally anti-symmetric, is invariant
under any bilateral rotations. This plays a role in the original
purification scheme \cite{Bennett}, in which arbitrary density
matrices are first mapped into a Werner state $x \left \vert \psi^-
\right \rangle \left \langle \psi^- \right \vert+\frac{1-x}{4}
\mbox{\boldmath$1$}$ \cite{Werner} without changing the weight of the
singlet state $f=\frac{1+3 x}{4}$ ($x$ is a real number) by bilateral
random rotations. The Werner state is diagonal in the Bell state basis
and with equal weight for all the elements except the singlet
state. Subsequently, Alice and Bob apply bilateral CNOT operations and
local measurements. By communicating and selecting a sub-ensemble of
the original ensemble of pairs they can distill a number of singlets.

However, for three (many) particles, there is no maximally entangled
state which is invariant under trilateral (multi-lateral) rotations
(for a classification of entangled states based on invariance under
local unitary transformations, see \cite{Linden}). This makes it more
difficult to transform an arbitrary state into Werner states.
This is why we cannot treat multi-particle entanglement
purification protocols as straightforward extensions of the
two-particle case.

In this letter, we propose {\it direct} purification protocols for a
wide range of mixed diagonal states having $N$-particle entanglement.
Our aim is to investigate the fidelity limits and efficiency for
purification and to make a first step towards a protocol that purifies
general mixed states. Our procedures have important implications for
the understanding of multi-particle entanglement and important
practical applications e.g., in quantum communications.  A central
result is that purifying multi-particle entangled states directly is
more efficient than relying on two-particle purification.

Although there is no maximally entangled state invariant under random
bilateral rotations for $N \geq 3$ ($N$ is the number of entangled
particles), we call the state
\begin{eqnarray}
\rho_W=x \left \vert \phi^+ \right \rangle \left \langle \phi^+ \right
\vert+\frac{1-x}{2^N} \mbox{\boldmath$1$}
\label{eqn:werner}
\end{eqnarray}
a ``Werner-type state'' because of the similarity with the two
particle case. Note that we write $\left \vert \phi^+ \right \rangle$
instead of $\left \vert \psi^- \right \rangle$ which has been done for
convenience.  The aim of purification is the distillation of a
sub-ensemble in the state $\left \vert \phi^+ \right \rangle$. The
fidelity, i.e. $\left \langle \phi^+ \right \vert \rho_W \left \vert
\phi^+ \right \rangle $, of the Werner-type state is
$f=x+\frac{1-x}{2^N}$.  These Werner-type states are practically
important for transmitting $N$ entangled particles to $N$ different
parties via noisy channels as we will explain later.

In the following, we present a protocol (P1+P2 in Fig~\ref{fig:pur}),
which can can purify a Werner-type state provided the fidelity of the
initial mixed state is higher than a certain critical value.  The
advantage of this protocol is that Werner-type states for {\it any
number of particles} can be {\it directly} purified.  
\begin{minipage}{3.27truein}
\begin{figure}
  \begin{center}
    \leavevmode
    \epsfxsize=3.27truein
    \epsfbox{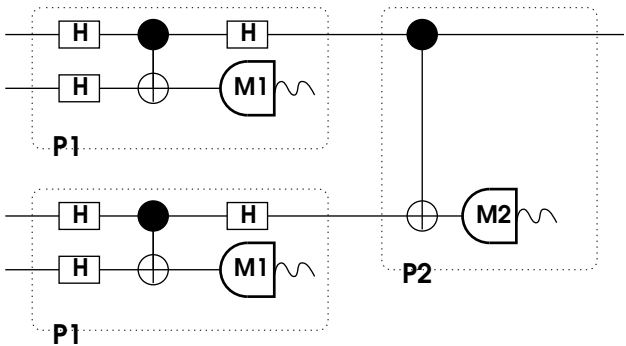}
  \end{center}
\caption{ 
Purification protocol P1+P2. H is a Hadamard transformation, M1 and
M2 are local measurement and classical communication. This diagram
shows four particles belonging to Alice.  Bob and others apply exactly
the same procedure.
}
\label{fig:pur}
\end{figure}
\smallskip
\end{minipage}

In the protocol P1+P2, each party (Alice, Bob and others) perform
iterations of the operations P1 followed by P2 on the particles
belonging to them.  The operation P1 consists of a local Hadamard
transformation which maps $\left \vert 0 \right \rangle \rightarrow
\left( \left \vert 0 \right \rangle + \left \vert 1 \right \rangle
\right)/ \sqrt{2}$, $\left \vert 1 \right \rangle \rightarrow \left(
\left \vert 0 \right \rangle - \left \vert 1 \right \rangle \right)/
\sqrt{2}$, a local CNOT (Control NOT) operation and a measurement M1,
and another local Hadamard transformation. In M1, we keep the control
qubits if an even number of target qubits are measured to be in the
state $\left \vert 1 \right \rangle$, otherwise the control qubits are
discarded.  For example when purifying for three particles, we only
keep $\left \vert 000 \right \rangle$, $\left \vert 011 \right
\rangle$, $\left \vert 101 \right \rangle$, $\left \vert 110 \right
\rangle$.  The operation P2 consists of a local CNOT operation and a
measurement M2 in which we keep the control qubits if all target bits
are measured to be in the same state, otherwise the control qubits are
discarded.  For example, when purifying three particles, we only keep
$\left \vert 000 \right \rangle$ and $\left \vert 111 \right \rangle$.
In this operation, the diagonal and off-diagonal elements of the
density matrix are independent of each other, so that the off-diagonal
elements do not affect the purification.

Our purification scheme is, however, not restricted to Werner-states.
When the state to be purified is $\left \vert \phi^+ \right \rangle$,
we call the state $\left \vert \phi^- \right \rangle$ the pairing
state of $\left \vert \phi^+ \right \rangle$.  If the initial mixed
state does not have any weight on the pairing state and weights on
other states are equal or some could be zero, iterations of the
operation P2 only is sufficient to purify the initial ensemble to the
$\left \vert \phi^+ \right \rangle$ state.  This purification
procedure fails if the weight of $\left \vert \phi^- \right \rangle$
is not exactly zero, because even a very small weight of $\left \vert
\phi^- \right \rangle$ in the initial mixed state results in an even
distribution of $\left \vert \phi^+ \right \rangle$ and $\left \vert
\phi^- \right \rangle$ after iteration and destroys entanglement.  The
reason for this is not entirely understood.  

When the initial state has weight only on the pairing states, that is,
when we have states of the form
\begin{eqnarray}
\rho=f \left \vert \phi^+ \right \rangle \left \langle \phi^+ \right \vert
+ \left(1-f \right)
\left \vert \phi^- \right \rangle \left \langle \phi^- \right \vert,
\label{eqn:snwnps}
\end{eqnarray}
then these can be purified by the iteration of only the operation
P1. P1 maps the state Eq.~(\ref{eqn:snwnps}) into a state of the same
form to Eq.~(\ref{eqn:snwnps}) but new fidelity $f^\prime=f^2/\left(2
f^2 - 2 f +1\right)$.  That is, the states with the initial fidelity
$f$ can be purified to $\left \vert \phi^+ \right \rangle$ if $f>1/2$
from the condition $f^\prime-f>0$.  For $f<1/2$, P1 purifies into
$\left \vert \phi^- \right \rangle$. When $f=1/2$, the resulting state
is disentangled and therefore cannot be purified by local operations
and classical communications.

In our purification protocols, we purify many-particle entangled
states {\it directly}.  This is necessary for fundamental
investigation of characteristic multi-particle entanglement.  However,
you could imagine schemes which purify many-particle entanglement via
two-particle purification: one of these schemes for three particles
(of Alice, Bob, and Claire) uses the fact that we know how to purify
two particles.  So this scheme converts three particle states into two
particle states, then purifies these two particle states, and finally
re-converts them to three particle entangled states.  This involves
the following: (1) Divide an ensemble of the state for three particles
into equal amount of two sub-ensembles.  (2) Bob measures his particle
from one sub-ensemble in the state $\left \vert \chi^\pm \right
\rangle=\left (\left \vert 0 \right \rangle \pm \left \vert 1 \right
\rangle \right)/\sqrt{2}$ and Claire measures her particle from
another sub-ensemble in the same state $\left \vert \chi^\pm \right
\rangle$. If Bob or Claire project onto $\left \vert \chi^- \right
\rangle$, then they inform Alice to perform the $\sigma_z$ operation,
so that the final two particle ensemble is in the same state as after
a projection onto $\left \vert \chi^+ \right \rangle$ after which
Alice does nothing.  Then we have two reduced two-particle entangled
states (one pair shared by Alice and Bob and another pair shared by
Alice and Claire).  (3) Perform the purification protocol
\cite{Bennett,Deutsch} to each of the entangled state of two
particles.  Then we get maximally entangled two particles shared
between Alice and Bob, and between Alice and Claire.  (4) Alice
performs a CNOT operation on her two particles, and then measures the
target particle.  Then we obtain the maximally entangled GHZ state
\cite{Zeilinger}.

We next analyse this scheme and compare it to our direct purification
schemes.  Any efficient direct three particle purification scheme {\it
should} perform better than this indirect method via two particles
because one obtains {\it one} maximally entangled state of three
particles from {\it two} maximally entangled states of two
particles. For purification of $N$-particle entangled states, we get
one maximally entangled state from $N-1$ maximally entangled states of
two particles.  In addition, the number of two-qubit CNOT operations,
each of which is difficult to carry out practically to high accuracy,
is higher than in our direct scheme. These ``inefficiencies'' are the
main practical disadvantage of the two particle scheme.  In the
following, we investigate the fidelity limit and efficiency of
purification and show that direct multi-particle purification is
indeed the more efficient method.

For two-particle entanglement, an initial fidelity $f>1/2$ is
sufficient for successful purification if we have no knowledge of this
initial state \cite{Bennett,Deutsch}.  However, the sufficiency
condition is not as simple for more than three particles.  We have
found several different criteria, depending on the type of mixed
states.

For the Werner-type states of the form $\rho_W=x \left \vert \phi^+
\right \rangle \left \langle \phi^+ \right \vert+\frac{1-x}{2^N}
\mbox{\boldmath$1$}$, and purification by the protocol P1 + P2, we
obtain numerically the results shown in Table~\ref{table:werner}.
\begin{minipage}{3.27truein}
\begin{table}
\begin{center}
\begin{tabular}{|c|c|c|c|}\hline
{$N$}&{A}&
{B}&{C}\\ \hline
{$2$}&{$f \geq 0.5395$}&{$f>1/2=0.5$}&{$f>1/2$}\\
{$3$}&{$f \geq 0.4073$}&{$f>5/12 \approx 0.4167$}&{unknown}\\
{$4$}&{$f \geq 0.313$}&{$f>3/8=0.375$}&{unknown}\\
{$5$}&{$f \geq 0.245$}&{$f>17/48 \approx 0.3542$}&{unknown}\\
{$6$}&{$f \geq 0.20$}&{$f>11/32 \approx 0.3438$}&{unknown}\\
\hline
\end{tabular}
\smallskip
\caption{A: Observed fidelity limit of initial states to be purified
for $N$ particles of the Werner-type states by the protocol P1 + P2,
B: Theoretical fidelity limit of the purification scheme via
two-particle purification, and C: the theoretical minimum sufficient
fidelity for purification.}
\label{table:werner}
\end{center}
\end{table}
\end{minipage}
\vspace{-0.7cm}

The theoretical fidelity limit for the Werner-type states $\rho_W$ of
the purification scheme via two-particle purification is determined by
the condition that the fidelity $f_r$ of the reduced two-particle
states should be $f_r>1/2$.  For example, for three particles, the
Werner state having initial fidelity $f=x+\left(1-x \right)/8$ is
reduced to a two-particle state after the measurement of Bob or Claire
$\rho_r=x \left \vert \phi^+ \right \rangle \left \langle \phi^+
\right \vert+\frac{1-x}{4} \mbox{\boldmath$1$}$.  The fidelity of the
reduced two-particle state is now $f_r= \left(1+ 6 f \right)/7$. For
four particles, we have $f_r=\left(1+ 4 f \right)/5$, for five
particles, $f_r=\left(7+24 f \right)/31$, for six particles,
$f_r=\left(5+16 f \right)/21$ and so on.  The general formula for the
fidelity limit of purification scheme via two particle purification is
$f > \left(2^{N-1}+1 \right)/\left(3 \cdot 2^{N-1} \right)$ where $N$
is the number of particles, which tends to $1/3$ as $N$ tends to
infinity.

We see from Table~\ref{table:werner},that the protocol P1 + P2 is not
optimal for two particles.  So it may not be optimal for $N > 2$.
However, for more than three particles, our observed fidelity limit is
lower than that obtained by route via two-particle purification.  In
general, the fidelity limit decreases as the number of particles
increases.  We can say that any Werner-type state whose fidelity
satisfies the bounds in column A is entangled.  In fact, any state
that can be {\it locally} converted into a Werner-type state
satisfying column A is also entangled.  However the final boundary
separating entangled and disentangled states is still unknown.

For the states having no weight on $\left \vert \phi^- \right \rangle
\left \langle \phi^- \right \vert$ and equal weight on all other
states except $\left \vert \phi^+ \right \rangle \left \langle \phi^+
\right \vert$, the fidelity limit of purification by the protocol P2
is $f>2^{-\left(N-1\right)}$. The fidelity limit obtained by the
purification scheme via two-particle purification is $2/5=0.4$ for the
three-particle case, $65/23 \approx 0.35846 $ for the four-particle
case, $125/377 \approx 0.328912$ for five-particle case and so on,
i.e.  worse than that in our protocols.  

We have seen that direct many particle purification can purify states
that {\it cannot} be purified via the two particle purification scheme
described before. This already suggests that multi-particle
purification is also more efficient in terms of the number of
maximally entangled states one obtains.

We define the efficiency of our protocol by the product of the
survival probability of the control qubit $P_J$ after $J$ iterations
of the protocol and $1/2^J$, which originates from the fact that the
entanglement of the target qubits is destroyed.  The product of the
normalisation for each iteration gives the probability $P_J$ that we
keep the entangled state after $J$ iterations of the purification
procedure. The number of iterations $J$ is chosen such that the
fidelity reaches unity with some a priori chosen accuracy.  

The protocol P1 also purifies an ensemble of a pure state $\left \vert
\Phi \right \rangle =a \left \vert 0 0 \cdot \cdot \cdot 0 \right
\rangle + b \left \vert 1 1 \cdot \cdot \cdot 1 \right \rangle$, where
$b=\sqrt{1-a^2}$, into a sub-ensemble of the maximally entangled pure
state $\left \vert \phi^+ \right \rangle$, that is the state with
$a=b=1/\sqrt{2}$ (we assume $a \leq b$ for convenience).  The
efficiency of the purification protocol P1 for the pure state $\left
\vert \Phi \right \rangle$ is {\it invariant} for entangled states of
any number of particles and coincides with the efficiency of the
purification scheme of Deutsch et al \cite{Deutsch} for a two-particle
pure state.  The efficiency of our purification protocol of the pure
state for $N$-particle entangled state is ${N-1}$ times better than
that of the purification scheme via two-particle purification
\cite{Bennett,Deutsch}.

We compare the efficiency of our purification protocol for the
Werner-type states and that of the purification scheme via
two-particle purification using the ``normalised'' efficiency. The
normalised efficiency is the product of survival provability $P_J$ of
the control qubit for our protocol, but is $P_J/(N-1)$ for the
purification scheme via two-particle purification.  The factor
$1/(N-1)$ originates from the fact that one $N$-particle maximally
entangled state is obtained from $N-1$ two-particle maximally
entangled states.  In Fig.~\ref{fig:eff-werner}, we show the numerical
result for the normalised efficiency of the two purification
procedures for three particles against the initial fidelity $f$.  Our
direct purification scheme performs approximately equally well for an
initial fidelity in the range of $0.4 < f \le 0.5$ but, clearly
performs better for higher initial fidelity.  We have made the same
comparison for four-particle and higher order entanglement and note
that the direct purification scheme is always more efficient than the
scheme via two-particle purification.  In fact, the difference in
normalised efficiency between the two schemes becomes even larger for
higher order entanglement. This is an important result because it
shows that it is more advantageous both in terms of resources (number
of CNOT operations) and normalised efficiency (number of maximally
entangled states obtained) to perform direct purification of our type
than to rely on the two-particle purification schemes.
\begin{minipage}{3.27truein}
\begin{figure}
  \begin{center}
    \leavevmode
    \epsfxsize=3.27truein
    \epsfbox{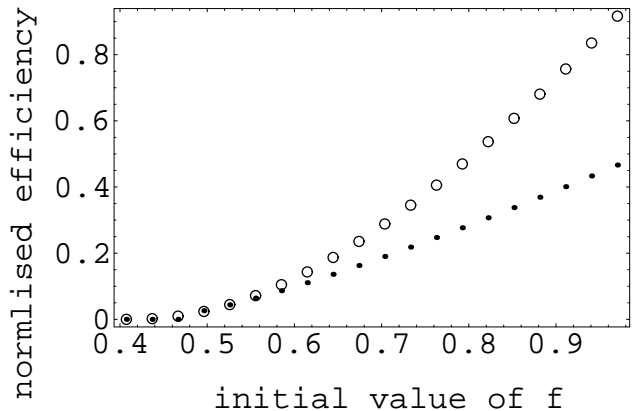}
  \end{center}
\caption{Normalised efficiency of purification of the Werner-type
states for three particles against the initial value of fidelity
$f$. The circles are obtained numerically by our purification protocol
P1 with a choice of accuracy $10^{-7}$.  The dots are obtained by the
purification scheme via two-particle purification with the same choice
of accuracy.}
\label{fig:eff-werner}
\end{figure}
\end{minipage}
\smallskip

Next we present an important example of a common noisy quantum
communication channel which gives rise to Werner-type states and where
our direct purification schemes can be successfully applied.  We show
that the mixed entangled states that we have treated in this letter
can be useful in practical applications. The mixed entangled states
are likely to appear when one has an ensemble of initially maximally
entangled states (for example, $\left \vert \phi^+\right \rangle$) of
$N$ particles and then transmits the $N$ particles to $N$ different
parties via noisy channels.  Let us consider the effect of a channel
whose action on each particle can be expressed by random rotations
about random directions. When each noisy channel causes random
rotations (about a random direction and by a random angle) with
probability $1-x$, while it leaves the particle unaffected with
probability $x$, the state after transmission becomes the Werner-type
state as in Eq.~(\ref{eqn:werner}).  If we consider a noisy channel
causing random rotations with a small but random probability depending
on the state, purification of states of high fidelity and small random
weights on other diagonal states will also be significant.  These
states are similar to Werner-type states but with additional random
weights on the diagonal elements.  When the ratio of the additional
random weight to fidelity is small, that is, the weight difference
among other diagonal elements are much smaller than the fidelity, we
have checked that the protocol P1 + P2 is successful.  However, the
final criterion for purification is not yet understood, as the success
of purification depends on the distribution of the diagonal elements.

We have found that combinations of the protocols P1 and P2 can
directly purify a wide range of mixed states of many particles.  The
advantage of the protocols proposed in this letter is that they can
{\it directly} purify some practically important states (Werner-type
states, states having no weight on the pairing state, etc.) of {\it
any number of particles}.  We have investigated the fidelity limit and
efficiency of the purification protocol and have shown that our direct
purification protocols are more efficient than two-particle schemes.

The fidelity limit of the initial states that are purifiable depends
on the distribution of the weight on other diagonal states.  This is a
condition of different character from the case of two particles
\cite{Deutsch}.  For two particles, the distribution of the weight on
other diagonal elements was irrelevant for purification,
since any distribution of weights on the other diagonal can be
transformed into an even distribution by local random rotations of
both particles without changing the amount of entanglement.  This
suggests that there may be some additional structure to entangled
mixed states for many particles, which does not exist for mixed
entangled states of two particles.

This work was supported in part by the UK Engineering and Physical
Sciences Research Council, the Knight Trust, the European Community,
the Alexander von Humboldt Foundation, and the Japan Society for the
Promotion of Science.

\end{multicols}


\vspace{-0.5cm}

\begin{thebibliography}{99}
\vspace{-1.5cm}
%
\bibitem{Barenco} A. Barenco, Contemporary Physics {\bf 37} 375 (1996).
%
\bibitem{Bennett0}
C.H. Bennett et al, Phys. Rev. Lett. {\bf 70} 1895 (1993).
%
\bibitem{Ekert}
A.K. Ekert, Phys.Rev.Lett. {\bf 68}  661 (1991).
%
\bibitem{Greenberger} D.M. Greenberger et al, Am. J. Phys. {\bf 58}
1131 (1990).
%
\bibitem{Bennett1} C.H. Bennett et al, Phys. Rev. A {\bf 54} 3824
(1996).
%
\bibitem{Horodecki} M. Horodecki et al, Phys. Rev. Lett. {\bf 78} 574
 (1997).
%
\bibitem{Vedral1} V. Vedral et al, Phys. Rev. Lett. {\bf 78} 2275
(1997).
%
\bibitem{Vedral2} V. Vedral et al, Phys. Rev. A {\bf 57} 4452 (1997).
%
\bibitem{Vedral3} V. Vedral and M.B. Plenio, Phys. Rev. A
{\bf 57} (1998).
%
\bibitem{Popescu2} S. Popescu and D. Rohrlich, Phys. Rev. A {\bf 56}
3219 (1997).
%
\bibitem{Bennett} C.H. Bennett et al, Phys. Rev. A {\bf 53} 2046
(1996) ; C.H. Bennett et al, Phys. Rev. Lett {\bf 76} 722 (1996).
%
\bibitem{Gisin} 
N. Gisin, Phys. Lett. {\bf 210} (1996) 151.
%
\bibitem{Deutsch} D. Deutsch et al, Phys. Rev. Lett. {\bf 77} 2818
(1996).
%
\bibitem{Werner}
R.F. Werner, Phys. Rev. A {\bf 40}, 4277 (1989).
%
\bibitem{Linden} N. Linden and S. Popescu, LANL e-print server,
quant-ph/9711016 (1997).
%
\bibitem{Zeilinger} A. Zeilinger et al, Phys. Rev. Lett. {\bf 78} 303
 (1997).
%
\end{thebibliography}
\end{document}